\documentclass[preprintnumbers, floatfix, showkeys, preprintnumbers, letterpaper, twocolumn, superscriptaddress,nofootinbib]{revtex4-2}
\pdfoutput=1 
\usepackage{graphicx}
\usepackage{microtype}
\usepackage{amsmath}
\usepackage{amssymb}
\usepackage{subfigure}
\usepackage{url}
\usepackage{hyperref}
\usepackage{mathtools}
\usepackage{orcidlink}
\usepackage{slashbox}
\usepackage{booktabs}
\usepackage{relsize}
\usepackage{xcolor}
\usepackage{color}
\usepackage{mathrsfs}
\usepackage{calrsfs}
\usepackage{amsfonts}
\usepackage{tabularx}
\usepackage{latexsym}
\usepackage{ragged2e}
\usepackage{epsfig}
\usepackage{textcomp}
\usepackage{float}

\usepackage{caption}
\DeclareCaptionJustification{justified}{\leftskip=0pt \rightskip=0pt \parfillskip=0pt plus 1fil}
\definecolor{vividviolet}{rgb}{0.62, 0.0, 1.0}
\definecolor{amaranth}{rgb}{0.9, 0.17, 0.31}
\definecolor{palatinateblue}{rgb}{0.15, 0.23, 0.89}
\definecolor{brightpink}{rgb}{1.0, 0.0, 0.5}
\definecolor{cornflowerblue}{rgb}{0.39, 0.58, 0.93}
\definecolor{deepcarminepink}{rgb}{0.94, 0.19, 0.22}
\definecolor{radicalred}{rgb}{1.0, 0.21, 0.37}

\hypersetup{ linktoc=all,
	colorlinks, linkcolor={palatinateblue},
	citecolor={brightpink}, urlcolor={amaranth}
}

\graphicspath{{Images/}}

\renewcommand{\d}[1]{\ensuremath{\operatorname{d}\!{#1}}}

\def\sideremark#1{\ifvmode\leavevmode\fi\vadjust{\vbox to0pt{\vss
			\hbox to 0pt{\hskip\hsize\hskip1em
				\vbox{\hsize1.3cm\tiny\raggedright\pretolerance10000
					\noindent #1\hfill}\hss}\vbox to8pt{\vfil}\vss}}}%
\def\beq{\begin{equation}}
\def\eeq{\end{equation}}

\begin{document}
\title{How Much Geometry Does Horizon Entropy Determine? \\A Discourse on Entropy, Gravity, and Observable Quantities}

\author{Yen Chin \surname{Ong}\orcidlink{0000-0002-3944-1693}}
\email{ongyenchin@nuaa.edu.cn}
\affiliation{Center for the Cross-disciplinary Research of Space Science and Quantum-technologies (CROSS-Q), College of Physics, Nanjing University of Aeronautics and Astronautics, \\29 Jiangjun Road, Nanjing City, Jiangsu Province 211106, China}

\begin{abstract}
Building on Jacobson's thermodynamics derivation of Einstein field equations, the ``generalized entropy and varying-$G$'' (GEVAG) framework was proposed to consistently incorporate generalized horizon entropy $S=f(A)/4G$ into both the thermodynamics and the geometry of spacetime. This leads to an effective gravitational coupling $G_\text{eff}=G/f'(A)$. A related ``entropy-gravity correspondence'' (EGC) instead is equivalent to promoting this horizon relation to a radially varying bulk coupling $G_\text{eff}(r)$. We clarify the similarities and fundamental distinctions between these approaches. Although they share the same horizon consistency relation, their global causal structures and asymptotic geometries can differ substantially. In addition, the radial prescription of EGC introduces an additional ambiguity for multi-center configurations. We further examine a near-horizon extension of GEVAG, which restores the equivalence between thermodynamic and ADM masses that is absent in the horizon-only formulation. We then discuss the implications of these different prescriptions for the Bekenstein bound. The Bekenstein bound provides a surprising link between $G_\text{eff}$ in near horizon GEVAG, EGC, and the one obtained in the recent $f(R)$-reconstruction approach of generalized entropy geometry via Wald entropy. 
Finally, we raise the most important question: what exactly is the $G$ measured in laboratories?
\end{abstract}

\maketitle
\section{Introduction: Generalized Entropy and Geometry}

In general relativity (GR), black holes have entropy given by the standard area law
\begin{equation}
S=\frac{A}{4G},
\end{equation}
where we have set $c=\hbar=k_B=1$. This is known as the Bekenstein-Hawking entropy. The fact that black hole entropy is an \emph{area law} has always been a puzzle. The remarkable thing may not merely be that entropy depends on area instead of the volume (given that ``black hole volume'' is rather nontrivial geometrically); it may be that in GR it is \emph{exactly linear} in area. 

In recent years, there has been much interest in generalizing the Bekenstein-Hawking entropy (hereinafter, ``generalized entropy''), not unlike the more familiar attempts at modifying gravity. Both approaches start with what holds in GR and try to change the ingredients to see what might happen to the theory. The hope is that this may lead to some possible explanations of dark energy or other yet unsolved problems in gravitational physics. At the very least, trying to change GR will help us to further appreciate the underlying structures of the theory. Some of the often used models include: Barrow entropy \cite{2004.09444},
Tsallis(-Cirto) entropy \cite{Tsallis,Tsallis2}, 
Kaniadakis entropy \cite{0210467,0507311,2109.09181}, 
and Sharma-Mittal entropy \cite{1802.07722,SM1,SM2}, as well as generalizations thereof \cite{2201.02424}. See the review part of \cite{2502.05801}.

In the earlier literature, usually generalized entropy is imposed as an additional ingredient on top of the GR solutions. That is to say, the geometry is unchanged, and only the entropy is modified. However, given the deep connection between thermodynamics and gravity, it is more natural to suspect that the change in the area entropy expression would also modify the gravity theory itself. Otherwise, inconsistencies may occur \cite{2109.05315,2207.07905}. This became rather concrete if one considers the Jacobson's approach \cite{9504004}, in which the Einstein field equations are derived from thermodynamic considerations. Following this idea, one could replace the area law that Jacobson used as an input of the derivation, by any entropy expression and try to develop the corresponding gravity theory. This results in the ``generalized entropy and varying-G'' (GEVAG) framework \cite{2407.00484}. Subsequently, this method was applied to various contexts, including the investigation of Bekenstein bound \cite{2505.03907} and cosmology \cite{2603.23551}. Its connection to generalized uncertainty principle (GUP) and quantum black holes were investigated in \cite{2505.07972,2604.01286}.

There is also another approach that is developed in parallel with GEVAG, which is now known as the ``entropy-geometry correspondence'' \cite{2511.04613,2512.13769} (hereinafter, EGC). Like GEVAG, it also tries to understand how generalized entropy may affect the spacetime geometry. EGC does not start from the Jacobson's approach, but instead use the standard first law of thermodynamics as a starting point. Given that generalized entropy is now actively progressing as a research field, it is important to further study and clarify how these theories are related.
We will review both approaches in the next section, before proceeding to analyze their similarities and differences.

We should also remark that the idea that a generalized horizon entropy should be accompanied by a corresponding modification of gravitational dynamics or black-hole geometry predates both GEVAG and EGC. For example, Asghari and Sheykhi \cite{2110.00059} derived modified gravitational field equations from Barrow entropy in September 2021, while Jusufi et al. \cite{2110.07258} subsequently reconstructed a Schwarzschild-like metric compatible with the temperature associated with Barrow entropy. In Ref.\cite{2207.09271}, proposed in July 2022, the relation between generalized entropy and modified gravity was further investigated, with an effective gravitational coupling obtained for Tsallis entropy that can be recognized as the special-case expression for GEVAG/EGC (but the results and interpretation for the Barrow case are somewhat inaccurate, or at least incomplete). Ref.\cite{2207.09271} also emphasized on the distinction between the thermodynamic energy entering the first law $\d E=T\d S$ and the mass parameter characterizing the spacetime (the ADM mass in the asymptotically flat case). Building upon these considerations, GEVAG was introduced in June 2024 \cite{2407.00484} as a general framework applicable to arbitrary entropy functions $S(A)=f(A)/4G$, with $G_\text{eff}=G/f'(A))$ providing the central entropy–gravity relation. Approximately sixteen months later, a parallel general construction was proposed in November 2025 \cite{2511.04613} and subsequently developed as a framework under the name entropy–geometry correspondence (EGC) \cite{2512.13769}. More recently, the relation between the entropy slope and the effective gravitational coupling has also emerged independently in the thermodynamic-gravity construction of Figliolia et al. \cite{2602.20430}, who obtained $G_\text{eff}=1/(4s_0)$ for an area-type entropy with constant slope $s_0$, thus providing an independent derivation of the effective-$G$ relation via a different route (Massieu-functional stationarity condition). Thus, while the broader idea that generalized entropy may modify gravitational dynamics or black-hole geometry has several antecedents and continues to arise from different thermodynamic approaches, GEVAG precedes EGC as a general framework relating an arbitrary horizon entropy to the corresponding modification of gravity.

The plan is as follows. In Sec.(\ref{s2}) we will review the constructions of GEVAG and EGC. This is followed by the analysis of the modified Schwarzschild solution in both theories. Specifically we will discuss the horizon and causal structure in Sec.(\ref{s3}), the thermodynamic mass in Sec.(\ref{s4}). We will see that GEVAG and EGC have a lot of similarities. In some sense, GEVAG proposed a modification to the black hole horizon, while EGC extends this to the whole space. However, this is an oversimplified statement as we will see that the two theories can give rise to very different causal structures. Then in Sec.(\ref{s5}) we will see that there is actually a hierarchy of theories one can consider between GEVAG and EGC, and in fact a near-horizon extension of GEVAG was already considered in the literature. However, now a closer inspection reveals that near-horizon extension of GEVAG enforces the identification between thermodynamic mass and ADM mass, which has been overlooked previously. Next, in Sec.(\ref{s6}), we examine what each theory has to say about the Bekenstein bound, a problem that plagues generalized entropy if $G$ is fixed. After that, we discuss the most important question from the point of view of observation in Sec.(\ref{s70}): what exactly do we mean by the ``Newton constant'' measured in laboratory, under these theories? Are there other possible observational signature?
We conclude with discussions in Sec.(\ref{s7}), focusing on the conceptual issues of both GEVAG and EGC, especially when we have more than one black holes to consider.

\section{The Construction of the Theories}\label{s2}

GEVAG is derived by following Jacobson's method \cite{9504004}. We first consider the matter flux
\begin{equation}
\delta Q = - \kappa \int_\mathcal{H} \lambda T_{ab} k^a k^b \text{d}\lambda \text{d}A
\end{equation}
through a horizon. This leads to
the change of the horizon area, given by
\begin{equation}
\delta A = - \int_\mathcal{H} \lambda R_{ab} k^a k^b \text{d}\lambda \text{d}A,
\end{equation}
in which $\lambda$ is an affine parameter, and $k^a$ the component of the horizon generator null vector. 

Using the Clausius relation $\delta Q=T\d S$, we then substitute the change in the entropy $\delta S=\frac{f'(A)}{4G} \delta A$ and
identify the temperature with the surface gravity $T=\kappa/2 \pi$. Finally we obtain
\begin{equation}
-\kappa \int T_{ab} k^a k^b \text{d}\lambda \text{d}A = - \frac{\kappa}{2\pi} \int \frac{f'(A)}{4G}  R_{ab} k^a k^b \text{d}\lambda \text{d}A.
\end{equation}
In order for this integral relation to hold for all null vectors $k^a$, the integrands must agree. In other words we must have
\begin{equation}
T_{ab} k^a k^b = \frac{f'(A)}{8\pi G} R_{ab} k^a k^b.
\end{equation}
Finally one obtains the field equations
\begin{equation}
R_{ab} - \frac{1}{2}R g_{ab} + \Lambda g_{ab} = 8\pi G_{\text{eff}} T_{ab},
\end{equation}
with $G$ replaced by 
\begin{equation}
G_\text{eff} = \frac{G}{f'(A)},
\end{equation}
assuming that $f'(A)\neq 0$.
This was referred to as the ``generalized entropy and varying-G'' (GEVAG) framework \cite{2407.00484}.
Equivalently,
\begin{equation}
G_\text{eff} = \frac{1}{4S'(A)}.
\end{equation}

On the other hand, EGC starts from assuming the metric function of a modified Schwarzschild metric
\begin{equation}
F(r) = 1 - Mg(r).
\end{equation}
This means that 
\begin{equation}
g(r_+) = 1/M.
\end{equation}

We consider a static spacetime so that $M$ is fixed, then
\begin{equation}
\frac{dF}{dr} = - M\frac{dg}{dr},
\end{equation}
so the Hawking temperature is\footnote{This also agrees with GEVAG identification. However, it is not clear if we can simply perform Wick rotation to identify the temperature with cyclic imaginary time as we did in GR, since that was inherited from statistical physics with the usual Boltzmann factor $e^{-\beta E}$. This may need to be modified if the black hole entropy is no longer the Bekenstein-Hawking one. This concern was already raised in \cite{2407.00484}. In fact, a generalization of this sort was explored afterward in \cite{2410.11044}.}
\begin{equation}
T=\left.\frac{F'(r)}{4\pi} \right|_{r=r_+}= -\left.\frac{Mg'(r)}{4\pi}\right|_{r=r_+}.
\end{equation}

In EGC, the standard first law of thermodynamics of black holes is assumed, i.e. 
\begin{equation}
\d M = T \d S.
\end{equation}
From this we immediately obtain
\begin{equation}\label{2}
\frac{\d M}{\d r_+} =T \frac{\d S}{\d r_+} = -\frac{M g'(r_+)}{4\pi} S'(r_+).
\end{equation}
From the horizon function definition, we also have $M(r_+)=1/g(r_+)$. This leads to
\begin{equation}\label{3}
\frac{\d M}{\d r_+} = -\frac{g'(r_+)}{g(r_+)^2}.
\end{equation}
Equating Eq.(\ref{2}) and Eq.(\ref{3}), we get the identification
\begin{equation}
-\frac{g'(r_+)}{g(r_+)^2} = -\frac{Mg'(r_+)S'(r_+)}{4\pi} = - \frac{g'(r_+)S'(r_+)}{4\pi g(r_+)},
\end{equation}
where for the second equality we again used the relation $M(r_+)=1/g(r_+)$. Assuming that $g'(r_+) \neq 0$, we get
\begin{equation}
\frac{1}{g(r_+)^2} = \frac{S'(r_+)}{4\pi g(r_+)}.
\end{equation}
Finally we obtain \emph{on the horizon}, 
\begin{equation}
g(r_+) = \frac{4\pi}{S'(r_+)}.
\end{equation}
EGC then \emph{promotes} this relation to
\begin{equation}
g(r) = \frac{4\pi}{S'(r)}, ~\forall r.
\end{equation}
Therefore the metric is proposed as 
\begin{equation}\label{FEGC}
F(r) = 1-\frac{4\pi M}{S'(r)}, 
\end{equation}
and therefore one obtains a modified geometry of the spacetime.

In the following, instead of using the function $S(r)$, we will instead for most of the time, refer to the quantity
\begin{equation}
G_\text{eff}(r) = \frac{2\pi r}{S'}.
\end{equation}
Not only does this make a direct comparison with GEVAG easier, it arguably provides a better ontology, since this language does not suggest that any constant $r$ surface literally possesses the generalized entropy $S(r)$. 
In fact, Wang and Braunstein \cite{2207.04390} have shown that an analogue of the first law generally fails for non-horizon surfaces.
Nevertheless, exact spherically symmetric spacetimes are exception, so the modified Schwarzschild case is strictly speaking not ruled out on this ground. Furthermore, as we shall see below EGC geometries need not be asymptotically flat, which is another assumption in \cite{2207.04390}. Nevertheless, \cite{2207.04390} suggests that one should be careful with extending an entropy of the horizon to non-horizon surfaces.
In \cite{2511.04613} $S(r)$ is given a possible interpretation as the vacuum entanglement entropy. However, it is not clear that this interpretation remains consistent for a generalized entropy. Entanglement entropy is conventionally defined by the von Neumann entropy. Under the usual typicality assumptions, the entanglement entropy of a sufficiently small subsystem is expected to approximately reproduce its thermodynamic entropy after the degrees of freedom of the larger subsystem are traced out. It is then unclear how a generalized thermodynamic entropy is to be recovered while retaining the standard von Neumann definition. One possibility is that the generalized entropy reflects a modified underlying state counting; alternatively, if the entropy functional itself is to be generalized, the corresponding notion of entanglement entropy, as well as the usual relation between entanglement and thermodynamic entropy, would have to be reconsidered. In other words, changing the thermodynamic entropy from $A/4G$ to $f(A)/4G$ is potentially saying something very nontrivial about the underlying state counting or its Hilbert-space structure. We are not yet ready to tackle this question in this work. Thus perhaps it is best to interpret EGC as an effective radially varying-$G$ theory\footnote{The mathematics remain unchanged, but the ontology would be rather different.}. 

The distinction is not merely terminological. Mathematical notation, like ordinary language, carries with it a conceptual picture. As Wittgenstein once said, ``a picture held us captive.'' Writing in terms of $G_\text{eff}$ instead of keeping it in terms of $S'(r)$ invites a different physical interpretation that consequently makes some physical questions more natural.

In a more recent paper Ref.\cite{2608.07046}, the authors tried to justify Eq.(\ref{FEGC}) as ``not an extrapolation of a horizon relation into the bulk'', but rather that ``every radius at which $S'$ is evaluated is an on-shell horizon radius of a different solution.'' This might have given a motivation for constructing $g(r)$ at every $r$ (or equivalently, $S'(r)$), but still the final metric is still a \emph{bona fide} extension \emph{within a single spacetime}, because for a fixed member of the family, the reconstructed $g(r)$ is evaluated also at radii that are not horizons of that spacetime and determines its geometry there. In other words, the fact that $g(r)$ can be reconstructed from horizon data across a family does not imply that there is no extension ``into the bulk''. Without such an extension, the metric would not differ from Schwarzschild geometry away from the horizon, which EGC geometries often do (see below). Still, there are some merits in the EGC approach, and it may be unnecessary at this stage to demand a deeper justification for the $g(r_+) \to g(r)$ prescription: one may simply regard it as an ansatz and investigate where it leads.

\section{Horizon and Geometry}\label{s3}

In GEVAG, because we have the field equation, we note that for the vacuum case, it is the same as GR. Therefore the Schwarzschild metric remains a solution, except that the gravitational coupling $G$ has to be replaced by $G_\text{eff}$. Thus the modified Schwarzschild metric function is simply
\begin{equation}
F_\text{GEVAG}= 1-\frac{2G_\text{eff}M}{r}.
\end{equation}
Given the exact form of the entropy, one can apply the area relation $A=4\pi r_+^2$ and solve $G_\text{eff}$ in terms of $G$, $M$, and the extra parameters in the specific generalized entropy.

On the other hand, the modified Schwarzschild metric in the EGC approach is
\begin{equation}
F_\text{EGC}= 1-\frac{4\pi M}{S'(r)}.
\end{equation}
At the horizon, we have 
\begin{equation}\label{4}
S'(r_+)=4\pi M. 
\end{equation}
Utilizing the chain rule, we have
\begin{equation}
S'(r) = \frac{\d S}{\d A}\frac{\d A}{\d r} = S'(A) 8\pi r.
\end{equation}
Thus we can write Eq.(\ref{4}) as
\begin{equation}
8 \pi r_+ S'(A_+) = 4\pi M.
\end{equation}
We still use prime to denote differentiation, the arguments would tell us with respect to with variable we are differentiating and thus there is no risk of confusion.

This means that 
\begin{equation}
r_+ = \frac{M}{2S'(A_+)}. 
\end{equation}
If we compare this to GEVAG equation,
\begin{equation}
r_+ = 2G_\text{eff}M = \frac{M}{2S'(A_+)},
\end{equation}
we see that they are exactly the same equation. 

However, this only means that the two theories share the same horizon equation, but their geometry overall can be very different. Essentially this can already be appreciated by noting that: GEVAG does not change the geometry away from the horizon, whereas EGC explicitly promotes the modified relation to all possible values of $r$. 

We can use an explicit example to illustrate this. Consider the power law form\footnote{This is essentially the Tsallis and Barrow forms.}
\begin{equation}
S(A)=\left(\frac{A}{4G}\right)^p.
\end{equation} 
In GEVAG, since
\begin{equation}
S=\frac{f(A)}{4G},
\end{equation}
we have
\begin{equation}
G_\text{eff}(A) = \frac{G}{f'(A)} = \frac{1}{4S'(A)}, 
\end{equation}
and thus quite simply
\begin{equation}
G_\text{eff}(A) = \frac{G}{p}\left(\frac{A}{4G}\right)^{1-p}.
\end{equation}
Substituting $A=4\pi r_+^2$, we can obtain finally the horizon as
\begin{equation}
r_+ = \left(\frac{2MG^p}{p\pi^{p-1}}\right)^{\frac{1}{2p-1}},
\end{equation}
and the metric function as
\begin{equation}
F_\text{GEVAG}=1-\frac{2GM}{r}\left[\frac{1}{p}\left(\frac{A}{4G}\right)^{1-p}\right].
\end{equation}

Now, for EGC, we have
\begin{equation}
S(A) = \left(\frac{A}{4G}\right)^p \Rightarrow S(r) = \left(\frac{\pi r^2}{G}\right)^p.
\end{equation}
The derivative of the promoted function $S(r)$ is
\begin{equation}
S'(r)= \frac{2p \pi^p}{G^p} r^{2p-1}.
\end{equation}
Therefore the metric function becomes
\begin{equation}
F_\text{EGC}=1-\frac{2MGp}{p\pi^{p-1} r^{2p-1}}=1-\left(\frac{r_+}{r}\right)^{2p-1}.
\end{equation}

Therefore we indeed have the same horizon, namely,
\begin{equation}
r^\text{GEVAG}_+ = r^\text{EGC}_+,
\end{equation}
but the geometry is different. In particular, in GEVAG the modified Schwarzschild geometry is always asymptotically flat. However, this is not true in EGC. In fact, to recover $F \to 1$ at large $r$, we would need to impose $p > 1/2$. Of course, asymptotic flatness is not just about the Minkowski limit, but also how fast the mass term decays, and so even for $p > 1/2$, the black hole ``mass'' is not quite the usual ADM mass.

Another interesting example is the generalized entropy of the form 
\begin{equation}\label{renyi}
S_R = \frac{1}{\alpha}\ln\left(1+\frac{\alpha A}{4 G}\right), ~\alpha>0,
\end{equation}
known as the R\'enyi entropy. 

The two theories agree on the location of the horizon.
However, the spacetime causal structure is very different.
In GEVAG, we note that
\begin{equation}\label{5}
G^\text{GEVAG}_\text{eff}=G\left(1+\frac{\alpha A}{4G}\right).
\end{equation}
Using the area-mass relation $A=16\pi G_\text{eff}^2M^2$, we get a quadratic equation in $G_\text{eff}$:
\begin{equation}
4\pi \alpha M^2 G_\text{eff}^2 -G_\text{eff}+G=0.
\end{equation}
This gives two possible values of $G_\text{eff}$:
\begin{equation}
G_\text{eff} =\frac{1\pm \sqrt{1-16\pi \alpha G M^2}}{8\pi \alpha M^2}.
\end{equation}
These in turn give two \emph{branches} for the solution. The GR branch corresponds to the negative sign in front of the square root. Namely, if we want to recover GR in the small correction limit, we should choose the solution with $G_\text{eff}^{(-)}$, as 
\begin{equation}
\lim_{\alpha \to 0} G_\text{eff}^{(-)} = G.
\end{equation}
The other branch gives a strong coupling in the small correction limit, since
\begin{equation}
G_\text{eff}^{(+)} \sim \frac{1}{4\pi \alpha M^2} \to \infty,
\end{equation}
as $\alpha \to 0$.

On the other hand, in EGC, we promote Eq.(\ref{5}) to
\begin{equation}
G^\text{EGC}_\text{eff}(r) =G\left(1+\frac{\pi \alpha r^2}{G}\right), \forall r.
\end{equation}
The metric function is then modified into
\begin{equation}\label{metricr}
F_\text{EGC}(r)=1-\frac{2M}{r}\left(G+\alpha \pi r^2\right) = 1-\frac{2GM}{r}-2\pi\alpha M r.
\end{equation}
This gives rise to \emph{two horizons}:
\begin{equation}
r_{\pm} = \frac{1\pm \sqrt{1-16\pi \alpha GM^2}}{4\pi \alpha M}.
\end{equation}
The inner horizon has the behavior 
\begin{equation}
r_- = 2GM + O(\alpha), 
\end{equation}
which recovers the GR Schwarzschild horizon as $\alpha\to 0$, whereas the outer horizon
\begin{equation}
r_+ \sim \frac{1}{2\pi \alpha M} \to \infty
\end{equation}
in the same limit.

Thus, in the case of R\'enyi entropy, we see that GEVAG interprets the two possible $G_\text{eff}$'s as two separate branches, each of which is an asymptotically flat geometry. On the other hand, EGC treats $G_\text{eff}$ as a function in a single spacetime, which gives rise to a non-asymptotically flat geometry with two horizons. In fact, the deviation from GR becomes increasingly large away from the horizon for EGC due to the term linear in $r$ in the metric function\footnote{The effects and interpretations of such a term, which can arise in various modified gravity theories, was studied in details in \cite{2106.05205}.}. This is not a surprise since if we expand the R\'enyi entropy as power series in $\alpha$, we note that
\begin{equation}
S_R = \frac{A}{4G} -\frac{\alpha}{2}\left(\frac{A}{4G}\right)^2 + \cdots, 
\end{equation}
which means for any fixed $\alpha$, no matter how small, there is always a sufficiently large $A$ such that the second term dominates over the first and ceases to be a ``small correction''. (A similar behavior motivates Ref.\cite{2205.09311} to consider possible ``running'' of the generalized entropy parameter.)

For completeness and easier reference, we shall provide the effective gravitational coupling as well as their asymptotic behaviors for the often-used generalized entropies, including the logarithmic correction found in quantum gravity literature, in Table \ref{entropies}. 
The $G_\text{eff}(A)$ is in the GEVAG language, but they are equivalent to the EGC results obtained in \cite{2511.04613}. The asymptotic behaviors are discussed in details in \cite{2608.07046}.
\begin{table}[t]
\centering
\footnotesize
\setlength{\tabcolsep}{2pt}
\begin{tabular}{lccc}
\toprule
Entropy & $S$ & $G_{\rm eff}$ & $|1-f_{\rm EGC}|$ \\
\midrule
\midrule
Tsallis--Cirto
& $\dfrac{A_0}{4G}y^\delta$
& $\dfrac{G}{\delta}y^{1-\delta}$
& $\sim r^{1-2\delta}$
\\
\cmidrule(lr){1-4}

R\'enyi
& $\dfrac{\ln(1+\alpha x)}{\alpha}$
& $G(1+\alpha x)$
& $\sim r$
\\
\cmidrule(lr){1-4}

Barrow
& $\dfrac{A_0}{4G}y^{1+\Delta/2}$
& $\dfrac{G}{1+\Delta/2}y^{-\Delta/2}$
& $\sim r^{-1-\Delta}$
\\
\cmidrule(lr){1-4}

Kaniadakis
& $\dfrac{\sinh(Kx)}{K}$
& $G~\text{sech}(Kx)$
& $\sim r^{-1}e^{-K\pi r^2/G}$
\\
\cmidrule(lr){1-4}

Log-Correction
& $x + C\ln(x)$
& $\dfrac{G}{1+\frac{C}{A}}$
& $\dfrac{2\pi GMr}{\pi r^2+CG}$
\\
\bottomrule
\end{tabular}
\caption{Generalized entropies and their effective couplings,
$G_{\rm eff}=(4S_A)^{-1}$. The last column gives the
large-$r$ behavior of the EGC metric. To make a compact table, we have defined
$x\equiv A/(4G)$ and $y\equiv A/A_0$, where $A_0$ is a constant. For
GEVAG, the geometry is always asymptotically flat.}
\label{entropies}
\end{table}

We note that the asymptotic behavior for the R\'enyi case is especially peculiar because the linear behavior diverges at large $r$. There is a way to appreciate why it is special. Consider the two-family generalization of Shamal-Mittal form. As we shall see, the Shamal-Mittal entropy provides an interesting observation to categorize the asymptotics of EGC.
Write $x \equiv A/4G$, then we have the Shamal-Mittal entropy\footnote{We use $\mu$ and $\nu$ as the deformation parameters instead of the original notations ($R$ and $\delta$) to avoid confusions, since we have used those letters for other quantities in this work.}
\begin{equation}
S_\text{SM}(A)=\frac{1}{\mu}\left[(1+\nu x)^{\mu/\nu}-1\right].
\end{equation}
Under GEVAG, we have
\begin{equation}
G^\text{GEVAG}_\text{eff}(A)=G(1+\nu x)^{1-\mu/\nu},
\end{equation}
which under EGC, is promoted to
\begin{equation}
G^\text{EGC}_\text{eff}(r)=G\left(1+\nu \frac{\pi r^2}{G}\right)^{1-\mu/\nu}.
\end{equation}
Asymptotically, the EGC geometry given by Eq.(\ref{metricr}) satisfies
\begin{equation}\label{mu}
|1-F_\text{EGC}(r)|\sim r^{1-2\mu/\nu}.
\end{equation}
This, if $\mu/\nu > 1/2$, we have $|1-F_\text{EGC}(r)| \to 0$, wheares for $\mu/\nu < 1/2$, we have a divergent behavior for $|1-F_\text{EGC}(r)|$. The case $\mu/\nu = 1/2$ gives a constant for $|1-F_\text{EGC}(r)|$. 
The R\'enyi limit of Shamal-Mittal entropy is when $\mu \to 0$, since
by noting that 
\begin{equation}
(1+\nu x)^{\mu/\nu} = \exp\left[\frac{\mu}{\nu}\ln(1+\nu x)\right],
\end{equation}
the small $\mu$ power series expansion for the exponential gives us 
\begin{flalign}
S_\text{SM}(A) &= \frac{1}{\mu}\left[(1+\nu x)^{\mu/\nu}-1\right]\\
&=\frac{1}{\mu}\left[1+\frac{\mu}{\nu}\ln(1+\nu x) + O(\mu^2)-1\right],
\end{flalign}
and thus
\begin{equation}
\lim_{\mu \to 0} S_\text{SM}(A) =\frac{1}{\nu}\ln(1+\nu x),
\end{equation}
which is exactly the R\'enyi entropy given in Eq.(\ref{renyi}).
Eq.(\ref{mu}) then gives $|1-f_\text{EGC}(r)| \to r$. R\'enyi entropy exhibiting growing linear metric function is due to the fact that it sits at the corner where the two-family parameter controlled by $\mu/\nu$ goes to zero. (Ordinary Bekenstein-Hawking entropy corresponds to $\mu/\nu=1$.) 

\section{Thermodynamic Mass}\label{s4}

While the EGC approach, the mass of the black hole enters the first law $\d M=T\d S$, in the GEVAG framework, it is the thermodynamic energy or thermodynamic mass $\d E=T \d S$; see also \cite{2106.00378}. That is, given the entropy expression, and the Hawking temperature
\begin{equation}
T=\frac{1}{8\pi G_\text{eff} M}=\frac{1}{2\sqrt{\pi A}},
\end{equation} 
one integrates the first law to get the energy expression. 
Explicitly, 
\begin{equation}
\d E = \frac{1}{2\sqrt{\pi A}}\frac{\text{d}A}{4G^\text{GEVAG}_\text{eff}(A)},
\end{equation} 
and thus
\begin{equation}\label{EHGEVAG}
E(A) = \frac{1}{8\sqrt{\pi}}\int_0^A \frac{f'(\tilde{A})}{G\sqrt{\tilde{A}}}\text{d}\tilde{A}.
\end{equation}
In general this can be rather complicated.

Consider again our toy model as an explicit example.
\begin{equation}
S=\left(\frac{A}{4G}\right)^p. 
\end{equation}
GEVAG prescription gives
\begin{equation}
G_\text{eff} = \frac{G}{p}\left(\frac{A}{4G}\right)^{1-p}. 
\end{equation}
The area-mass relation $A=4\pi(2G_\text{eff}M)^2$ gives, after some algebraic manipulation,
\begin{equation}
A=M^{\frac{2}{2p-1}}\left[4\pi^{\frac{1}{2}}\frac{G}{p}\left(\frac{1}{4G}\right)^{1-p}\right]^{\frac{2}{2p-1}}.
\end{equation}
The first law is 
\begin{flalign}
\d E &= \frac{1}{8\pi G_\text{eff} M} \frac{p}{4G}\left(\frac{A}{4G}\right)^{p-1}\text{d}A \\
&=\frac{1}{8\pi}\left(\frac{p^2}{4G^2}\right)\left(\frac{A}{4G}\right)^{2p-2} \frac{1}{M}\text{d}A.
\end{flalign}
We can then re-write the mass $M$ in terms of $G$ and $A$, and integrate. After which, we can re-write $A$ in terms of $M$ again to finally obtain a simple result:
\begin{equation}\label{energy}
E=\frac{M}{2p-1},~~p\neq \frac{1}{2}.
\end{equation}
Thus in this case the thermodynamic energy is just a constant multiple of the ADM mass. For the energy expressions for the R\'enyi case, see \cite{2505.03907}.

Given that the toy model formally has the Tsallis-Cirto form, this is essentially the result obtained in \cite{2407.00484}.
However, precisely because of this, along with the area-dependent $G_\text{eff}$, one can avoid the Bekenstein bound violation problem, which we will discuss in Sec.(\ref{s6}).

\section{Hierarchy of Theories}\label{s5}

We have seen that GEVAG only relates the generalized entropy to the effective gravitational constant at the horizon. The field equation itself does not tell us what to do elsewhere. On the other hand, EGC assumes outright that we should promote the effective $G$ to all possible values of $r$. In other words, the modified metric function in two theories can be directly compared as
\begin{equation}\label{gevagf}
F_\text{GEVAG}(r)=1-\frac{2MG_\text{eff}(A(r_+))}{r},
\end{equation}
and
\begin{equation}\label{egcf}
F_\text{EGC}(r)=1-\frac{2MG_\text{eff}(A(r))}{r}, ~\forall~\text{possible}~ r,
\end{equation}
respectively. That is to say, EGC interprets the entropy correction as a ``radially running'' Newton coupling.
We emphasize that despite the similarity in the \emph{form} of the equations above, the solutions can have very different spacetime geometry and causals structures, as discussed in the preceding sections. The reason is easy to see: in EGC, $G_\text{eff}(A(r))$ itself is a function of $r$, so when expanded out explicitly it need not be asymptotically flat.

In any case, Eq.(\ref{gevagf}) and Eq.(\ref{egcf}) suggest a possible hierarchy between the two theories: if GEVAG only restricts to the horizon, and EGC considers all possible value of $r$, can we consider something in the middle? The answer is yes, and surprisingly, are already considered in the literature from both the GEVAG and EGC sides.

GEVAG has a variant in which $G_\text{eff}$, which was initially only defined on the horizon, is extended into an $\varepsilon$-neighborhood of the horizon. To distinguish between these two versions, we will refer to the second one as ``near-horizon GEVAG'' or NH-GEVAG, while the original one can be emphasized as ``horizon GEVAG'', or H-GEVAG. The near-horizon version has the virtue that it could obtain a zero temperature black hole remnant in the small mass limit towards the end of Hawking evaporation \cite{2604.01286}, whereas H-GEVAG leads to a finite temperature ``remnant'' of the type first discovered in the GUP literature \cite{0106080}, which was argued to be somewhat inconsistent \cite{2604.01286}. 

Explicitly, given the entropy $S=f(A)/4G$, the Hawking temperature is:
\begin{equation}
T_\text{H-GEVAG} = \frac{1}{8\pi G_\text{eff} M},
\end{equation}
and
\begin{equation}\label{TNH}
T_\text{NH-GEVAG} = \frac{1}{8\pi G_\text{eff} M}+ 2r_+\frac{f''(A)}{f'(A)}.
\end{equation}
It can be directly checked that EGC agrees with the latter:
\begin{equation}
T_\text{EGC} = T_\text{NH-GEVAG}.
\end{equation}

For our power law example $S =(A/4G)^p$, we get\footnote{For example, see \cite{anand} for the Hawking temperature of the Barrow case in a more general setting. To compare with our model $S\propto A^p$, set the charge $Q$ to zero and the cosmological constant length scale $l \to \infty$ in the results therein, and use the fact that $1+\Delta/2=p$ where $\Delta$ is the Barrow parameter.}
\begin{equation}
T_\text{EGC} = T_\text{NH-GEVAG}=\frac{2p-1}{4\pi r_+}.
\end{equation}
Here we see why $p \geqslant 1/2$ is important, because it enforces the non-negativity of the Hawking temperature. (See \cite{2608.07046} for the recent EGC discussion on the positive temperature conditions.)
Interestingly, in the H-GEVAG case, we also require $p \geqslant 1/2$ if we impose\footnote{This was discussed already in \cite{2207.09271}.} $E \geqslant 0$; see Eq.(\ref{energy}). Note, however, that the $p=1/2$ case corresponds to formal divergent energy and vanishing temperature and should probably be avoided on that ground. 

In fact, there is one unexpected consequence for allowing a near-horizon extension of GEVAG. In my previous paper \cite{2604.01286}, the thermodynamic energy of a black hole in NH-GEVAG was derived from $\d E =T \d S$ and found to be\footnote{In \cite{2604.01286}, I noticed that although NH-GEVAG reproduced GUP temperature exactly, but the log-correction term in the entropy expression obtained from GUP differs by a minus sign with the one assumed to derive the GEVAG metric. In that paper I suspected that this may be due to the fact that $E\neq M$, but since we have found that $E=M$ in NH-GEVAG after all, this explanation fails. However, in GUP approach the horizon is also still the GR horizon, so perhaps we should not worry too much about its inconsistency given its heuristic nature.}
\begin{equation}\label{E}
E(A) =\frac{\sqrt{A}}{4}\frac{f'(A)}{\sqrt{\pi}G},
\end{equation}
c.f. Eq.(\ref{EHGEVAG}) for H-GEVAG.
The calculation is short enough for us to simply reproduce here. In NH-GEVAG, the Hawking temperature, Eq.(\ref{TNH}) can also be written explicitly in terms of the area as
\begin{equation}
T = \frac{1}{2\sqrt{\pi A}} + \sqrt{\frac{A}{\pi}}\frac{f''(A)}{f'(A)}.
\end{equation}
The first law is then
\begin{flalign}
\d E &= \frac{1}{4\sqrt{\pi}G}\left[\frac{f'(A)}{2\sqrt{A}} + \sqrt{A} f''(A)\right]\d A\\
&=\frac{1}{4\sqrt{\pi}G}\frac{\text{d}(\sqrt{A}f'(A))}{\text{d} A}\d A.
\end{flalign}
The remarkable fact that the terms in the square bracket is a total derivative is what allows the equation to easily integrate into Eq.(\ref{E}).

Since in H-GEVAG, we have $E \neq M$, I did not realize then that this is actually identically equal to the ADM mass $M$. To see this, we start from 
\begin{equation}
r_+ = 2G_\text{eff}M =\frac{2GM}{f'(A)},
\end{equation}
and thus we observe that
\begin{equation}\label{M}
M =\frac{r_+ f'(A)}{2G}.
\end{equation}
But the area-radius relation is
\begin{equation}
r_+ = \sqrt{\frac{A}{4\pi}}=\frac{\sqrt{A}}{2\sqrt{\pi}},
\end{equation}
and therefore Eq.(\ref{M}) becomes \emph{exactly} the same as Eq.(\ref{E}).
Thus, thermodynamic energy is identically the same as the ADM mass once we define $G_\text{eff}$ away from the horizon, even in just its infinitesimal neighborhood! This is worth emphasizing:
\begin{equation}
E_\text{H-GEVAG} \neq M_\text{H-GEVAG};
\end{equation}
\begin{equation}
E_\text{NH-GEVAG} \equiv M_\text{H-GEVAG};
\end{equation}
\begin{equation}
E_\text{EGC} \equiv M_\text{EGC}.
\end{equation}
The last one automatically holds since the EGC approach assumes from the beginning that $\d M=T\d S$.
 
From the EGC side, it has also been suggested that at least for some choice of the entropy, we need not promote $G^\text{EGC}_\text{eff}(r)$ to all values of $r$, but rather to some cutoff value $r=r^*$, beyond which a ``large-distance completion'' may be needed \cite{2608.07046}. This was suggested for the R\'enyi entropy discussed earlier precisely because of the linear term in the metric function. This geometry is not asymptotically flat and the authors in \cite{2608.07046} suggested that perhaps it should be interpreted as an effective infrared geometry valid only over a finite radial domain, or else supplied with a suitable large-distance completion. We will return to this point later.

\section{Bekenstein Bound}\label{s6}

Let us now discuss the Bekenstein bound, which has been analyzed in the GEVAG context in a previous paper \cite{2505.03907}. 
It is useful to define the dimensionless quantity
\begin{equation}
B = \frac{2\pi RE}{S},
\end{equation}
so that the Bekenstein bound
\begin{equation}
S\leqslant 2\pi RE
\end{equation}
becomes $B \geqslant 1$.

For fixed $G$, and if $E=M$, one notices that 
\begin{equation}
2\pi RE= 4\pi GM^2 = \frac{A}{4G},
\end{equation}
so that
\begin{equation}
B = \frac{A}{4GS} \geqslant 1
\end{equation}
translates to 
\begin{equation}
S \leqslant \frac{A}{4G}.
\end{equation}
This is exactly how one can interpret the Bekenstein bound: given a fixed size and fixed energy, the system cannot have entropy more than that of the Bekenstein-Hawking entropy of a black hole.

The issue is that, as was previously pointed out in the literature, the Bekenstein bound can be grossly violated for generalized entropy if we still use the GR horizon $R=r_+=2GM$ with fixed $G$, together with the mass $M$ as the the energy \cite{2207.13652, 2411.00694}. This can be appreciated from the expression
\begin{equation}
B = \frac{A}{4GS}.
\end{equation}
For, say, $S\propto A^p$ as in our previous model, clearly $B$ becomes area-dependent, and it can become less than unity for sufficiently large black holes. This is rather puzzling if we insist that large black holes should be well-described by standard GR black hole thermodynamics. 

For H-GEVAG, the Bekenstein bound has been studied in details in a previous paper \cite{2505.03907}. The energy expression is obtained from the first law $\d E = T \d S$, where $R=\sqrt{A/4\pi}$. This gives
\begin{equation}
\d E = \frac{1}{4\pi R} S'(A) \d A = \frac{S'(A)}{2\sqrt{\pi A}}\d A.
\end{equation}
Integrating yields the H-GEVAG energy
\begin{equation}
E(A)= \frac{1}{2\sqrt{\pi}}\int_0^A \frac{S'(\tilde{A})}{\sqrt{\tilde{A}}} \text{d}\tilde{A},
\end{equation}
which is just the same equation we obtained before in Eq.(\ref{EHGEVAG}) written in terms of the entropy.
This consequently implies
\begin{equation}
B = \frac{\sqrt{A}}{2S(A)} \int_0^A \frac{S'(\tilde{A})}{\sqrt{\tilde{A}}} \text{d} \tilde{A}.
\end{equation}

The NH-GEVAG version is simpler because, as we have established in the previous section, $E\equiv M$.
We first recall that
\begin{equation}
M = \frac{R}{2G_\text{eff}} = 2RS'(A),
\end{equation}
since $G_\text{eff}=1/4S'(A)$. 
The RHS in the Bekenstein is thus
\begin{equation}
2\pi RE = 2\pi RM = 4\pi R^2 S'(A) = AS'(A).
\end{equation}
Consequently we have a rather neat relation:
\begin{equation}\label{B}
B = \frac{AS'(A)}{S(A)} = \frac{\mathrm{d} \ln S}{\mathrm{d} \ln A}.
\end{equation}
The EGC expression is the same as NH-GEVAG's. It should be noted that this combination was suggestively interpreted  as a RG-scaling dimension
of the entropy functional $f(A)$, or equivalent of $S(A)$, in \cite{2604.01286}. This was also later noted in the EGC literature (Eq.(IV.17) of \cite{2608.07046}.)

We summarize the results in Table \ref{bekenstein-general}. We note that H-GEVAG is the only one that requires an integral expression. Recall that when we extend H-GEVAG to NH-GEVAG by extending the horizon to its $\varepsilon$-neighborhood, its thermodynamic energy changed from $E\neq M$ to $E \equiv M$. Physically, the $\varepsilon$-shell provides an additional constitutive information about the geometry (instead of energy in a direct manner -- since one does not expect energy in the  $\varepsilon$-shell to be sufficient to cause a finite change to $E$), and changes the temperature by a finite amount, so that the integral over the black hole states gives a finite shift to enforce $E \equiv M$.  

\begin{table}[t]
\centering
\small
\setlength{\tabcolsep}{3.5pt}
\begin{tabular}{lccc}
\toprule
Framework & $E$ & $B$ & $B\geqslant 1$ \\
\midrule
\midrule
Fixed $G$
& $M$
& $\dfrac{A}{4GS}$
& $S\leqslant\dfrac{A}{4G}$ \\
\cmidrule(lr){1-4}
H-GEVAG
& $\dfrac{\mathcal{I}(A)}{2\sqrt{\pi}}$
& $\dfrac{\sqrt{A}\,\mathcal{I}(A)}{2S}$
& $\dfrac{\sqrt{A}}{2}\mathcal{I}(A)\geqslant S$ \\
\cmidrule(lr){1-4}
NH-GEVAG
& $M$
& $\dfrac{A S'(A)}{S}$
& $A S'(A)\geqslant S$ \\
\cmidrule(lr){1-4}
EGC
& $M$
& $\dfrac{A S'(A)}{S}$
& $A S'(A)\geqslant S$ \\
\bottomrule
\end{tabular}
\caption{Comparison for a general entropy $S=S(A)$, where
$B\equiv 2\pi R E/S$ and
$\mathcal{I}(A)\equiv\displaystyle\int^A
(S_{\tilde A}/\sqrt{\tilde A})\,\text{d}\tilde A$.
The standard Bekenstein bound is $B\geqslant 1$.  The EGC result for the Bekenstein bound should be understood as only a formal statement; see the main text for details.}\label{bekenstein-general}
\end{table}

The results for EGC is only formal, in the sense that we derived the equations following the formula, but one has to be careful about whether the Bekenstein bound even holds or requires further modification in non-asymptotically flat spacetimes (since the EGC geometry can have different asymptotics). For example, in de Sitter space, the so-called ``D-bound'' becomes the more appropriate notion \cite{0012052}.

If we apply the general results in Table \ref{bekenstein-general} to the toy model $S(A)=x^p$, where $x \equiv A/4G$, then we arrive at Table \ref{bekenstein-special}. In this case, it is clear that the fixed $G$ version violates Bekenstein bound. 
Specifically, if $p>1$ then Bekenstein bound is satisfied only for $0<x<1$, namely for sub-Planckian black hole! Since one expects quantum gravitational corrections to be important for small black holes, we should restrict our discussion to $x \gg 1$.
In this case, the Bekenstein bound requires $p<1$. We also note that the EGC case is the same (formally) as NH-GEVAG, as expected given our previous discussions.
What is more interesting is the comparison between H-GEVAG and NH-GEVAG. The former gives $B=p/(2p-1)$, and so the Bekenstein bound holds if $1/2 \leqslant p \leqslant 1$. Thus the bound is violated for $p > 1$. However, just by extending off-horizon by an $\varepsilon$ amount, NH-GEVAG changes $B$ to just $p$, and so $p < 1$ is precisely when the bound is violated. In other words, the range of $p$ for which the Bekenstein bound is violated is exactly the opposite for H-GEVAG and NH-GEVAG! 

\begin{table}[h]
\centering
\small
\setlength{\tabcolsep}{3.5pt}
\begin{tabular}{lccc}
\toprule
Framework & $E$ & $B$ & $B\geqslant 1$ \\
\midrule
\midrule
Fixed $G$
& $M$
& $x^{1-p},x\gg 1$
& $p< 1$ \\
\cmidrule(lr){1-4}
H-GEVAG
& $\dfrac{M}{2p-1}$
& $\dfrac{p}{2p-1}$
& $\dfrac12<p\leqslant 1$ \\
\cmidrule(lr){1-4}
NH-GEVAG
& $M$
& $p$
& $p\geqslant 1$ \\
\cmidrule(lr){1-4}
EGC
& $M$
& $p$
& $p\geqslant 1$ \\
\bottomrule
\end{tabular}
\caption{Comparison for the toy model $S=x^p$, where $x\equiv A/(4G)$ and
$B\equiv 2\pi R E/S$. The standard Bekenstein bound is $B\geqslant 1$. 
For the fixed $G$ case we require large black holes and hence $x \gg 1$.
The EGC result for the Bekenstein bound should be understood as only a formal statement; see the main text for details.}\label{bekenstein-special}
\end{table}

We also note that for this specific example, the weak form of Bekenstein bound (namely that the RHS of Bekenstein bound is relaxed to $CRE$ with $C$ not necessarily $2\pi$, but still a fixed number), as proposed in \cite{2505.03907}, always holds for both forms of GEVAG and EGC, while it is violated for the fixed $G$ case. 

This is the appropriate place to discuss the recent proposal that one could construct an $f(R)$ theory from generalized entropy \cite{2608.25722}. The construction essentially relies on
\begin{equation}\label{wald} 
f_R \equiv \frac{\d f(R)}{\d R} = \frac{4G S(A)}{A}.
\end{equation}
Since the LHS is a function of the scalar curvature $R$, while the RHS is a function of $A$, this prescription requires one to choose a map $R(A)$, which is not unique. For example, the Schwarzschild-de Sitter geometry is a constant curvature solution whose metric function is
\begin{equation}
F(r) = 1- \frac{2GM}{r} - \frac{Rr^2}{12},
\end{equation}
since $R=4\Lambda$ in 4-dimensions. Solving for the horizon radius and thus the area, we have $A=A(R,M)$. Thus one has to make a choice as to what $M$ is, e.g. $M=0$ yields $A(R)=48\pi/R$. This is such a required map. In general, when the spacetime does not have a constant curvature, this approach would require more thoughts. For our purpose, it is sufficient to note that the construction then follows by applying the Wald entropy formula for constant curvature case:
\begin{equation}
S=\frac{A}{4G}f_R (R),
\end{equation}
and thus Eq.(\ref{wald}). 
In this approach, the natural coupling associated with this reconstruction is 
\begin{equation}\label{gfr}
G_\text{eff}^{f(R)} = \frac{G}{f_R} = \frac{A}{4S(A)}.
\end{equation}
This therefore differs from GEVAG/EGC assignment of
\begin{equation}
G_\text{eff}^\text{GEVAG/EGC} = \frac{1}{4S'(A)}.
\end{equation}
The difference is essentially that Eq.(\ref{gfr}) is equivalent to saying $S=A/4G_\text{eff}$, which is \emph{not} assumed in either EGC or GEVAG. In fact, we can now see that 
\begin{equation}
 G_\text{eff}^{f(R)}=G_\text{eff}^\text{GEVAG/EGC} \Longleftrightarrow AS'(A) = S(A).
\end{equation}
This is only satisfied for a linear law $S \propto A$. 
What is even more intriguing is that for NH-GEVAG and EGC, the ratio $AS'(A)/S(A)$ is exactly the ratio $B$ that we defined above in Eq.(\ref{B}) and Table \ref{bekenstein-general}. In other words, the ratio can be interpreted in 3 ways:
\begin{equation}\label{3ways}
B = \frac{AS'(A)}{S(A)} = \frac{\mathrm{d} \ln S(A)}{\mathrm{d} \ln A} = \frac{ G_\text{eff}^{f(R)}}{G_\text{eff}^\text{NH-GEVAG/EGC}}.
\end{equation}
The first equality follows directly from the definition of Bekenstein bound equality, the second one suggests a local scaling exponent interpretation, and the last is the ratio of the effective gravitational couplings between the Wald entropy approach and the EGC or NH-GEVAG approach.

In \cite{2608.07046} the authors introduced the ``anomalous exponent''
\begin{equation}\label{ano}
\eta_G(r) \equiv \frac{\mathrm{d}\ln G_\text{eff}^\text{EGC}(r)}{\mathrm{d}\ln r}.
\end{equation}
In terms of $\eta_G$, we have  
\begin{equation}
B= \frac{ G_\text{eff}^{f(R)}}{G_\text{eff}^\text{NH-GEVAG/EGC}}=1-\frac{\eta_G}{2}.
\end{equation}

It would be interesting to further investigate if there is a deeper reason why the Bekenstein bound ratio also measures the mismatch between the Wald entropy approach and the EGC or NH-GEVAG approach.

It should be emphasized that this is \emph{not} saying that GEVAG/EGC is not compatible with Wald entropy. If we already have an $f(R)$ gravity, then in the context of that theory, the natural identification is
\begin{equation}
G^{f(R)}_\text{eff} = \frac{G}{f_R}.
\end{equation}
The Wald entropy form can then be written as
\begin{equation}
S_\text{Wald} = \frac{A}{4G}f_R = \frac{A}{4G^{f(R)}_\text{eff}}.
\end{equation}
But it requires an additional assumption to say that given any generalized entropy $S(A)$, we can then write
\begin{equation}
G_\text{eff}= \frac{A}{4S(A)}.
\end{equation}
In fact as we have mentioned, this differs from GEVAG/EGC original identification
\begin{equation}
G_\text{eff}= \frac{A}{4S'(A)}.
\end{equation}
In fact, in \cite{2602.20430}, Figliolia et al. independently arrive at precisely the latter result in their local thermodynamic construction. Specifically, for an area-type entropy with locally constant slope $s_0$, they obtained 
\begin{equation}
G_\text{eff}= \frac{A}{4s_0}.
\end{equation}
In that work they also discussed how $f(R)$ gravity may arise. This was essentially the construction of Eling-Guedens-Jacobson \cite{0602001}, which established that curvature-dependent entropy densities supplemented by an internal entropy-production term gives $f(R)$ theory. Essentially, the integral form of Wald entropy (over the horizon $\mathcal{H}$) is
\begin{equation}
S_\text{Wald} = \frac{1}{4G}\int_\mathcal{H} f_R (R) \d A,
\end{equation}
where $f_R (R)$ can therefore be identified roughly as the entropy density $s(R)$.
The difference from \cite{2608.25722} is that in the latter, $S(A)$ is \emph{given}, and the question is to find an $f(R)$ theory whose Wald entropy agrees with $S(A)$. This has the issue of choosing the map $A(R)$, whereas the former, starting from $s(R)$, there is no such issue (everything is a function of $R$ throughout); though another ambiguity arise since a coarse-graining scale has to be fixed. We will return to this point in the Discussion section. 

For now let us return to why the Wald entropy need not be inconsistent with the GEVAG or EGC results. Brustein, Gorbonos and Hadad \cite{0712.3206} showed more generally that Wald entropy can be expressed as
\begin{equation}
S_W = \frac{A}{4G_W},
\end{equation}
where $G_W$ has a \emph{specific} interpretation, namely that it is associated with the kinetic coupling of the appropriate graviton polarization at the horizon. Thus, $G_W$ need not coincide with the coupling inferred from what essentially is a differential entropy change $\d S/\d A$. In fact, if 
\begin{equation}\label{wald2}
S_W(A) = \frac{A}{4G_W(A)},
\end{equation} 
then
\begin{equation}
\frac{1}{G_\text{GEVAG}} =4\frac{\text{d} S}{\text{d} A} = \frac{1}{G_W}\left(1-\frac{\mathrm{d} \ln G_W}{\mathrm{d} \ln A}\right).
\end{equation}
To see this, simply differentiate Eq.(\ref{wald2}) to get
\begin{equation}
\frac{\text{d}S}{\text{d}A} = \frac{1}{4G_W}\left(1-\frac{A}{G_W}\frac{\text{d}G_W}{\text{d}A}\right),
\end{equation}
and use the mathematical identity
\begin{equation}
\frac{A}{G_W}\frac{\text{d}G_W}{\text{d}A} = \frac{\mathrm{d}\ln G_W}{\mathrm{d} \ln A}.
\end{equation}
One can check that $G_W = G_\text{GEVAG}$ if $\mathrm{d}\ln G_W/\mathrm{d}\ln A = 0$, which is equivalent to the expressions in Eq.(\ref{3ways}) becoming unity. To see this, simply take the logarithm of Eq.(\ref{wald2}) and differentiate with respect to $\ln A$ to get
\begin{equation}
B = 1 -\frac{\mathrm{d} \ln G_W}{\mathrm{d}\ln A}.
\end{equation}

These discussions point us to the next important issue: there can be various distinct notions of ``gravitational coupling'' that in GR are equivalent, but in general need not be. Depending on what kind of operational procedures or experiments, we may define different kind of gravitational coupling instead of just ``the'' effective Newton constant.

\section{What do we actually observe?}\label{s70}

Once the metric is modified, we always have to be very careful about what we can really measure. For example, if a modified Schwarzschild metric function is of the form
\begin{equation}
1-\frac{2G\mathcal{F}(M,\eta)}{r}
\end{equation}
for some new parameter $\eta$, then it is tempting to fix $M$, and try to constrain the new parameter $\eta$. However, in practice the two quantities $M$ and $\eta$ are often not easily disentangled. Astronomers only see an effective mass $G\mathcal{F}(M,\eta)$. To fix $M$ is to have a ``God's eye view'', useful for theorists who want to study the effect of $\eta$ but not for observations to constrain the value of the parameter. Things can get complicated very fast when even $G$ is no longer a constant. In this section we aim to discuss how one may try to measure such an effect.

We first note that EGC rewrites the metric as
\begin{equation}
F(r) = 1-\frac{2MG^\text{EGC}_\text{eff}(r)}{r}.
\end{equation}
In the weak-field limit, like in GR, we should expect the gravitational potential
\begin{equation}
\Phi(r) = -\frac{MG^\text{EGC}_\text{eff}(r)}{r}.
\end{equation}
For a test mass $m$, the radial force in the Newtonian language would therefore be
\begin{equation}
F=-m\frac{\d \Phi}{\d r}.
\end{equation}
The derivative of the potential has two terms due to the radial running of the gravitational coupling. Explicitly,
\begin{flalign}
\frac{\d \Phi}{\d r} &= - M\frac{\text{d}}{\text{d} r} \left(\frac{G^\text{EGC}_\text{eff}(r)}{r}\right)\\
&=-M\left(\frac{{G'}^\text{EGC}_\text{eff}(r)}{r}-\frac{G^\text{EGC}_\text{eff}(r)}{r^2}\right).
\end{flalign}
Therefore 
\begin{equation}
F=-\frac{Mm}{r^2}\left(G^\text{EGC}_\text{eff}(r)-r{G'}^\text{EGC}_\text{eff}(r)\right).
\end{equation}
An experimentalist who performs a local experiment would infer the gravitational coupling
\begin{equation}
G_\text{force}(r)\equiv G^\text{EGC}_\text{eff}(r)-r{G'}^\text{EGC}_\text{eff}(r).
\end{equation}

We can check that this is consistent with the EGC's force expression in the literature. Start with 
\begin{equation}
G^\text{EGC}_\text{eff}(r) = \frac{2\pi r}{S'(r)},
\end{equation}
we differentiate to get
\begin{equation}
{G'}^\text{EGC}_\text{eff}(r) = \frac{2\pi}{S'(r)}-\frac{2\pi r S''(r)}{[S'(r)]^2}.
\end{equation}
This gives
\begin{equation}
G_\text{force}(r)=\frac{2\pi r^2 S''(r)}{[S'(r)]^2},
\end{equation}
as the two terms linear in $r$ canceled. 
Therefore, we obtain
\begin{equation}
F =  -\frac{Mm}{r^2}G_\text{force} = -2\pi mM \frac{S''}{(S')^2}.
\end{equation}

This is precisely the EGC-force law given in the foundational paper \cite{2511.04613}.
Since gravitational force is the gradient of the gravitational potential, it is sensitive to the local radial variation of the effective gravitational coupling, and hence the derive of $G^\text{EGC}_\text{eff}(r)$ matters. In other words, a local experiment designed to probe the inverse-square law does not only see $G^\text{EGC}_\text{eff}(r)$. Of course, one has to be very careful about what we mean by local experiment. A terrestrial experiment may not be applicable because the theory is constructed for a black hole spacetime, so one should instead probe the gravitational field around a black hole.  Existing EGC studies have already demonstrated that the radial continuation modifies photon spheres and black hole shadows \cite{2512.13769}, it is conceivable that future observations can further probe whether one could possibly detect the radially-varying term $r{G'}^\text{EGC}_\text{eff}$. For example, the angular frequency of a timelike circular geodesic is given by
\begin{equation}
\Omega=\sqrt{\frac{f'(r)}{2r}}=\sqrt{\frac{M}{r^3}\left[G^\text{EGC}_\text{eff}(r)-r{G'}^\text{EGC}_\text{eff}(r)\right]},
\end{equation}
where $f$ is the metric function
\begin{equation}
F(r)=1-\frac{2MG^\text{EGC}_\text{eff}(r)}{r}.
\end{equation}
That is,
\begin{equation}
G_\text{force}(r)=\frac{r^3 \Omega^2}{M},
\end{equation}
which can in principle be probed in astronomical observation, say from the orbital motion of a star around a black hole. 
Equivalently, instead of $\Omega$, we can use the period $T=2\pi/\Omega$, which suggests the use of the (modified) Kepler law. Indeed,
deviations from Keplerian orbital motion have long been employed as probes of modified gravity. See \cite{0606197,0612056,1603.03243} as examples. 
Better still, given the photon orbit and EHT constraint closer to the horizon, the stellar orbital dynamics (along with say, ISCO and accretion data) further out should help to test the EGC predicted radial profile of the effective gravitational coupling. 

There is an instructive historical precedent for taking such operational questions seriously. The remarkable empirical success of Newton's inverse-square law itself placed severe constraints on attempts to provide gravity with an underlying mechanical explanation. In considering an aether permeating space, Newton had to confront the question of why such a medium did not produce appreciable resistance to planetary motion. Among the possibilities he entertained was that material bodies might be sufficiently porous for the aether to pass through them. The historical details aside\footnote{Newton eventually gave up on a physical property of the aether and declared \emph{vim penetrantem spiritus}! See \cite{Dobbs}.}, the methodological point remains relevant to us today: a proposed gravitational ingredient must account not only for the effects attributed to it, but also for the \emph{absence} of effects that might otherwise be expected. Likewise, once $G_\text{eff}$ is promoted from a horizon quantity to a position-dependent gravitational coupling, one must ask what coupling would actually be inferred from an experiment, if any. If such an effect is not observed, one must eventually understand whether it is truly absent, suppressed, or simply beyond current observational sensitivity. Before asking whether an effect is detectable, the theory must tell us what the relevant observable is. For generalized entropy, this question needs to be investigated more extensively.

A more fundamental question, however, concerns what is the ``Newton constant'' $G$ that has already been measured in the lab. In the GEVAG picture, the proposal is that it is associated with the $G^\text{GEVAG}_\text{eff}$ of the cosmological horizon \cite{2407.00484,companion}. In the EGC case, this is somewhat unclear. The recent EGC paper, Ref.\cite{2608.07046}, refers to $G$ as the ``infrared Newton's constant'' but did not elaborate on it. One might be tempted to say that $G$ is the asymptotic value, in the sense that $G^\text{EGC}_\text{eff} \to G$ as $r \to 0$. However, this is not enough, as we should also require that the varying term $r{G'}^\text{EGC}_\text{eff}$ also goes to zero. In fact, instead of $r{G'}^\text{EGC}_\text{eff}$, one may want to use the ``anomalous exponent'' of Eq.(\ref{ano}) as the natural quantity that measures the ``running'' of $G^\text{EGC}_\text{eff}$, since
\begin{equation}
\eta_G(r) \equiv \frac{\mathrm{d}\ln G_\text{eff}^\text{EGC}(r)}{\mathrm{d}\ln r} = \frac{r{G'}^\text{EGC}_\text{eff}(r)}{G^\text{EGC}_\text{eff}(r)},
\end{equation}
and in addition this quantity is dimensionless and insensitive to the overall normalization of $G^\text{EGC}_\text{eff}$.

However, we have seen that the asymptotic geometries of EGC need not be flat, and the resulting $G^\text{EGC}_\text{eff}$ can be divergent or zero in that limit. 
For example, the Tsallis-Cirto and Barrow forms, which is schematically our toy model $S\sim A^p$, give rise to $G^\text{EGC}_\text{eff} \propto r^{2(1-p)}$ at large $r$ for $p>1/2$, and hence asymptoctically vanishes instead of approaching $G$. 
Thus taking the limit $r \to \infty$ may be problematic\footnote{The entropy with logarithmic correction term, $S(A)=A/(4G) + C \ln(A/4G)$, commonly derived from quantum gravity theories or models, is well-behaved in this sense. Its large $A$ behavior is $S = 1/(4G) + O(1/A)$, its effective gravitational coupling \cite{2407.00484,2603.23551,2505.07972,2604.01286} $G_\text{eff}$ also recovers $G$ in the large $A$ limit (or large $r$ in the ECG picture \cite{2608.07046}), and the running $\eta_G \to 0$.}. Perhaps it is more sensible to define the IR or Newtonian regime \emph{locally} over some radial interval by requiring that 
\begin{equation}
\zeta_G \equiv \left\lvert\frac{G^\text{EGC}_\text{eff}-G}{G^\text{EGC}_\text{eff}}\right\rvert \ll 1, ~~ |\eta_G|\ll 1.
\end{equation}
The R\'enyi case, for example, can then have a meaningful intermediate region that one can study. However, since $\zeta_G$ would eventually become large far away from the black hole, this still does not tell us how an observer far away (like us) should interpret our measured Newton constant. This is why the call of Ref.\cite{2608.07046} for an ``infrared completion'' becomes important, or even necessary. Thus for the R\'{e}nyi case we see very clearly that just requiring that $\zeta_G \sim 0$ and $\eta_G \sim 0$ is not sufficient, one needs a complete global prescription for the geometry that is sensible, that is \emph{not} the naive extension of EGC. 

It is worth emphasizing that there are various possible ``gravitational couplings'' here in EGC: the one constructed initially $G^\text{EGC}_\text{eff}$, the one measured by local experiment around a black hole $G_\text{force}$, and the one measured locally far away from the black hole, say, on Earth. Gravitational wave could provide another testing arena. In EGC, because we do not yet have the theory to deal with two black holes, binary black hole system cannot be used. However, the ringdown phase could provide some observable signature that shows deviation from GR. However, likewise $G^\text{EGC}_\text{GW}$ could be another quantity altogether. 
Such distinctions can be found in many modified theories of gravity. For example,
in Horndeski theory, there are various types of gravitational couplings that need not coincide \cite{2212.09094}.
As another, explicit example, consider the DHOST theory, in which $G_\text{GW}=1/(16\pi f)$, where $f=f(\phi)$ is a function of the scalar field. On the other hand, exterior to a gravitating source, one instead has \cite{2104.02445}
\begin{equation}
G_\text{ext} = \frac{1}{16\pi f\left(1-X\frac{f_X}{f}\right)},
\end{equation} 
where $X\equiv -\phi^a\phi_a/2$.
Historically such distinctions have always been important. For example, Clifford Will made exactly such a distinction in the context of Brans-Dicke theory, between what he call $G_{\infty}$, the overall cosmological background gravitational coupling measured far away from matter distribution, and the locally measured $G$ in the lab \cite{Will}. The extra complication in our case comes from black holes, which give rise to a different gravitational coupling. 

We leave another important question for the Discussion section: the Universe does not consist of only one black hole. Given the multitudes of black holes and each of them gives rise to a $G^\text{EGC}_\text{eff}$, how does one interpret a locally measured Newton constant?

\section{Discussion: Much Ado About Centers}\label{s7}

Let us first summarize the main results.
How gravity can arise from entropy has been investigated from numerous perspectives in the past. In addition to Jacobson's work, there was also the program of Padmanabhan \cite{0911.5004,0912.3165}, who tried to derive gravity from horizon thermodynamics. Verlinde \cite{1001.0785} also proposed the possibility of deriving gravity via an ``entropic force'', in which information stored on a ``holographic screen'' is utilized. 
The approach of Bianconi, on the other hand, starts from a geometric relative entropy to obtain a gravitational action and field equation \cite{2408.14391,2510.22545}. 
In these approaches, gravity is derived from thermodynamics or even more fundamentally quantum information, i.e. gravity is emergent. However, in this work we are concerned about another problem: if the entropy of a black hole horizon is modified, what is the effect on the gravity theory and spacetime geometry itself. In principle this question can be asked whether or not gravity is emergent.

We have discussed two main current approaches to understand the deep connections between gravity and entropy: the GEVAG approach that starts from the Jacobson method, and the EGC approach that starts from the first law of black hole thermodynamics. 
Central to both is the message that one cannot simply impose generalized entropy on GR solution because generalizing the horizon entropy would change the corresponding gravity theory, and therefore the underlying geometry. The question is how much of the geometry can be determined once the entropy function is given. The answer depends on what further assumptions we are willing to make.

Essentially, both approaches give rise to a new effective gravitational coupling $G_\text{eff}$. They disagree on how far away from the horizon should $G_\text{eff}$ be allowed to extend. Thus, starting from 
\begin{equation}
G_\text{eff}(A) = \frac{1}{4S'(A)},
\end{equation}
they all agree that this holds on the horizon, but EGC makes the extra assumption that the form can be promoted over a finite or global radial domain, and hence obtain a radial profile $G_\text{eff}(r)$. 
There also exists a near-horizon extension for GEVAG (``NH-GEVAG'') that allows the extension over an $\varepsilon$-neighborhood of the horizon. This version requires $G_\text{eff}$ on the horizon, and its first derivative there. 
In other words, NH-GEVAG requires only the local germ\footnote{In mathematical terminology, a \emph{germ} at a point represents the local behavior of a function near that point: two functions define the same germ if they agree in some neighborhood of the point, irrespective of how they differ farther away. Here we use the term to emphasize that NH-GEVAG requires only near-horizon information, and makes no commitment concerning a global radial continuation. EGC \emph{assumes} that the extended function is the one obtained by simply replacing $r_+$ by $r$, although mathematically there are infinitely many functions that will agree with the horizon value but differ elsewhere (similar assumption of function extension was also considered in constructing GUP effective metric \cite{2303.10719}).} of $G_\text{eff}(r)$ at the horizon rather than a globally defined radial profile.
We note that such an infinitesimal extension already renders the thermodynamics considerably cleaner (e.g. thermodynamic mass is identical to ADM mass). We should emphasize again that, EGC geometry is \emph{not} simply obtained by extending GEVAG prescription to finite values of $r$ or even globally, since although the horizon relation is the same in both theories, the solutions can have different causal structures. Notably for the R\'enyi entropy case, GEVAG has two possible $G_\text{eff}$'s, each correspond to a single-horizon spacetime, but in EGC, there are two horizons in a single spacetime.

For the discussion, it is instructive to ask what is the ontological status of $G_\text{eff}$.
An interesting conceptual issue arises when we go beyond a single black hole for the EGC case. 
With a single black hole, it is clear what\footnote{Henceforth, we drop the superscript ``EGC'' to prevent cluttering; it is clear we are talking about EGC.} $G_\text{eff}(r)$ means: an observer at some point $x$ located at coordinate distance $r$ away from the black hole experiences\footnote{Remember the caveats from the preceding section that experiments do not always probe $G_\text{eff}$ cleanly. However, for the sake of argument let us assume that we have a way to do so.} locally, the gravitational coupling $G_\text{eff}(r)$, or rather the combination $G_\text{eff}(r) M$.  When there are multiple black holes, the situation is less clear. Take the case of two identical mass black holes for example. Consider an observer located at distance $r_1$ from the first black hole and distance $r_2$ from the second black hole, the value of  $G_1\equiv G_\text{eff}(r_1)$ need not be the same as $G_2\equiv G_\text{eff}(r_2)$. If the observer had thought that $G$ is fixed, then said observer would simply assign two different masses to the black holes, so there is no conceptual problem as yet. However, consider a \emph{different} observer, located at distance $r'_1$ away from the first black hole and distance $r'_2$ away from the second black hole. Then assuming fixed GR would lead to the second observer assigning mass $G_1'$ and $G_2'$ to the black hole, respectively. So if different observers were to communicate and compare notes, they will either suspect that $G$ is not fixed, or that mass is observer-dependent! 

One may ask what happens if instead of the observational level, we move to a linear or weak-field regime of the underlying gravitational theory. Then the gravitational potential can be written (following GR), 
\begin{equation}
\Phi(x) = -\sum_i \frac{M_i G_\text{eff}(|x-x_i|)}{|x-x_i|},
\end{equation}   
for the multi-black hole case. Then $G_\text{eff}$ can be interpreted as a source-observer response kernel. Thus at the same spacetime point, it is perfectly fine to have $G_\text{eff}(x,x_1) \neq G_\text{eff}(x,x_2)$. 
In \cite{2608.07046}, similar to the mass-assignment discussion above, the authors reinterpret a point mass as a smeared effective source:
\begin{equation}
M\delta^{(3)}(r) \longmapsto MK(r).
\end{equation}
Thus an observer assuming GR might say that the observed mass 
\begin{equation}
M_\text{obs} = \frac{G_\text{eff}}{G}M \equiv u(r) M.
\end{equation}
This can then be interpreted as an \emph{effective mass}
\begin{equation}
M_\text{obs} = M_\text{eff} = 4\pi \int_0^r r'^2 \rho_\text{eff}(r') \d r'
\end{equation}
with 
\begin{equation}
\rho_\text{eff}(r) = MK(r).
\end{equation}
In other words, the re-interpretation \emph{keeps $G$ fixed but dresses the source}.

An operator that takes a delta source into an extended profile can be obtained as
\begin{equation}
K(r) = \mathcal{A}^{-2}(-\nabla^2)\delta^{(3)}(r).
\end{equation}
Fourier transforming this gives
\begin{equation}
\tilde{K}(k) = \mathcal{A}^{-2} (k^2),
\end{equation}
which for a generic entropy function would not be a finite polynomial in $k^2$. This implies that the object 
\begin{equation}
\mathcal{A}^{-2}(-\nabla^2)
\end{equation}
is a nonlocal operator instead of a finite-order differential operator\footnote{This follows from Peetre's Theorem in harmonic analysis.}.
The authors in  
\cite{2608.07046} thus wrote a nonlocal kernel $K(x,y)$, so that
\begin{equation}
\Phi \sim \int d^3y K(x,y)\rho(y),
\end{equation}
where say, for the two black hole case, where the black holes are located at $y=x_1, x_2$, the mass density is
\begin{equation}
\rho_\text{eff}(y) = M_1 K(|y-x_1|) + M_2 K(|y-x_2|).
\end{equation}
Then, the gravitational potential is
\begin{equation}\label{K}
\Phi(x) \sim M_1 K(x,x_1) + M_2 K(x,x_2).
\end{equation}
At the linear level this construction makes sense.
The problem is that we do not yet have a consistent rule for how multiple-source responses can be combined at a fully nonlinear level, where a simple superposition is no longer guaranteed to hold. More generally, as emphasized in \cite{2608.07046} regarding the definition of a nonlocal gravitational coupling, ``uniqueness can only be established after specifying the complete covariant action.'' 

Of course, one could also, in principle, consider $G_\text{eff}(r)$ as a field. However, for this one needs an equation of motion (EOM), c.f. the EOM for the scalar field $\phi$ (which takes the role of $1/G$) in the Brans-Dicke theory. EGC does not yet have a field equation of this form. 

On the other hand, GEVAG has a field equation, yet it still also lacks an EOM for $G_\text{eff}$. This is because in GEVAG, $G_\text{eff}$ is not a field. 
Furthermore, since GEVAG does not extend $G_\text{eff}$ to a finite radius away from the horizon, 
its most natural interpretation is not as a response to a source, but a gravitational response or \emph{susceptibility} due to the changes in the entropy, as recently proposed in a companion paper \cite{companion}:
\begin{equation}
G_\text{eff}^i \sim \left(4\frac{\partial S}{\partial A_i}\right)^{-1},
\end{equation}
where $i$ labels the individual black hole in the multi-black hole case.

Curiously, these discussions have a historical precedent of sort. We mentioned in Foonote 8 that Newton gave up on giving a physical property for the aether. However, he did contemplate that the aether could be distributed in a nonuniform manner. He specifically asked in Query 21 of \emph{Opticks} whether \cite{opticks}
\begin{quote}
``Is not this Medium much rarer within the dense Bodies of the Sun, Stars, Planets and Comets, than in the empty celestial Spaces between them? And in passing from them to great distances, doth it not grow denser and denser perpetually, and thereby cause the gravity of those great Bodies towards one another, and of their parts towards the Bodies; every Body endeavouring to go from the denser parts of the Medium towards the rarer?'' 
\end{quote}
In modern language, Newton was essentially considering the density of $\rho_\text{aether}=\rho_\text{aether}(r,M)$, where $M$ is the mass of a gravitating body. In fact he went one step further and asked if the gradient of the aether field is responsible for producing gravity. This idea has been developed many times since by various authors\footnote{For example, Thomas Young described material bodies as surrounded by aetherial ``atmospheres'' \cite{Cantor}.}. What is relevant to us is this: consider two massive bodies $M_1$ and $M_2$, what can we say about $\rho_\text{aether}$? Namely, how do two aether profiles combine? This is exactly the kind of ``multi-center problem'' we have been discussing. In fact, in the work of Mossotti in 1837 \cite{mossotti}, we found something intriguing. Mossotti wrote down the equilibrium aether density as, in schematic form (and modern notation)\footnote{Note that the mathematical form $e^{-ar}/r$ is that of a Yukawa profile, though here it describes density and not a potential. He could write down terms like $\sum_i e^{-ar}/r$ because the Yukawa form is a solution to a Green's function (modified Helmholtz equation with a point source), and Green's functions can be superposed.}
\begin{equation}
\rho_\text{aether}(x) = \rho_0 + \mathlarger{\sum}_{i}C_i\frac{e^{-ar_i}}{r_i},
\end{equation}
where $a$ is a constant and $\rho_0$ the background aether density. In other words there is a sum of perturbation terms on the aether background 
\begin{equation}
\rho_\text{aether}(x) = \rho_0 + \sum_{i}\delta_\text{aether}\rho(|x-x_i|),
\end{equation}
which resembles our previously discussed linear prescription of the gravitational coupling\footnote{Massive scalar-tensor theories can give rise to effective gravitational coupling of the form $G(1+K e^{-m r})$, where $K$ is some constant \cite{Järv,2305.06752}, thus the gravitational potential would take precisely the Yukawa form mentioned above! Another famous example is massive gravity at the linear level \cite{1105.3735}.}
\begin{equation}
G^\text{EGC}_\text{eff}(x) = G + \sum_i \delta G_i(|x-x_i|),
\end{equation}
or equivalently in terms of the effective smeared source description \`a la Eq.(\ref{K}). (The difference is of course that the aetherial atmosphere is supposed to physically there, but the latter only represents a nonlocal modification of the gravitational sector.)

Our problem is that we now know gravity to be a nonlinear phenomenon, and so the full theory cannot be describe as such a linear superposition. 
On the other hand, Oliver Lodge had insisted on the uniform nature of the aether. To explain gravity, he proposed instead a new property of the aether that he referred to as the aetherial ``tension'' or stress, $T=T(x)$. He called this a ``condition or state of the medium'' \cite{oliver}, whose variation gives rise to gravitational force. Namely, $T(r) \propto M/r$ and $\d T/\d r \propto M/r^2$. While this tension looks just like the modern gravitational field, we should note that the important interpretation here is that in Lodge's view, the underlying fundamental aether is universal, but the state associated with the aether is not. The comparison our effective gravitational coupling is then as follows: $G^\text{GEVAG}_\text{eff}$ inherited from the cosmological horizon is universal, but it can nevertheless exhibits a source-dependent local response on a black hole horizon. The gravitational susceptibility interpretation of GEVAG proposed in \cite{companion}
\begin{equation}\label{sus}
G^\text{GEVAG}_\text{eff} = \frac{1}{4}\chi^{-1}_\text{g}(A), \text{where}~\chi_g(A)\equiv\frac{\partial S}{\partial A},
\end{equation} 
is especially apt for such a description, which in turn also suggest how one may further consider ``coupling'' between various horizons. 

In the recent $f(R)$ gravity approach \cite{2608.25722} to generalized entropy, the construction is different from both GEVAG and EGC. Using the Wald entropy approach, one first chooses a branch that relates the area with the curvature scalar $R$. In this approach, since there \emph{is} a field equation, the geometry itself would determine $R(x)$ at any point $x$ of the spacetime. Thus, in terms of the effective gravitational coupling, we would have essentially
\begin{equation}
G_\text{eff}(x) \sim \frac{G}{f_R[R(x)]},
\end{equation}
where $f_R \equiv \d f(R)/\d R$. 
However, the ambiguity in that approach, as emphasized by the authors, is that the choice of the map $A(R)$ is not unique, and there is no clear principle of choosing one over the other in general. 
Once the action has been reconstructed, the theory gives $R(x)$ as a covariant scalar and so there is no ``multi-center'' problem to solve, the problem now occurs at the earlier stage, namely which family of horizons should we use to establish the correspondence between $A$ and $R$. 

The approach of \cite{2602.20430} is almost opposite to that of \cite{2608.25722}. Figliolia et al. starts with a local entropy density $s_0=\frac{\partial S}{\partial A}\rvert_{A_*}$ to get $G_\text{eff}(A_*)$. However, the problem is then the coarse-graining area $A_*$. This is arbitrary unless there is some principle to fix it. For this, the authors introduced their ``topological calibration principle''. 

These two constructions therefore illustrate complementary aspects of the inverse problem: the fundamental issue is that $S(A)$ alone does not uniquely prescribe how its thermodynamic information is to be promoted to bulk gravitational dynamics. In the $f(R)$ reconstruction of  \cite{2608.25722} the missing information is encoded in the nonunique map $R(A)$, whereas in the local thermodynamic construction of \cite{2602.20430} it appears as the choice of coarse-graining scale $A_*$, for which an independent calibration principle must be provided. Once the reference scale $A_*$ is fixed, however, a field equation\footnote{Once $A_*$ is specified, $s_0$ is held fixed under the local geometric variation used to derive the field equations. The resulting field equations is essentially the same as GEVAG's, but instead of $G_\text{eff}=G_\text{eff}(A)$ being a function of the horizon area, here instead $G_\text{eff}=1/4s_0$, and $s_0=s_0(A_*)=\frac{\partial S}{\partial A}\rvert_{A_*}$ is \emph{locally fixed}, i.e. during the variation one has $\delta S =s_0 \delta A$. The local patch therefore does not have to continually ``know'' the total horizon area. In spirit this is closer to Jacobson's original approach; whereas in GEVAG, the construction is not strictly local due to the area integral for nonlinear $S(A)$ does not allow area-dependence to drop out. Roughly speaking, the information about the nonlinearity of $S(A)$ in the approach of \cite{2602.20430} went into the calibration condition, and the topological calibration principle is then needed to connect the locally constructed $s_0$ with the ``global'' generalized entropy. Therefore we see that a linear entropy law conceals the role of ``locality'' in Jacobson's derivation of GR. When $S(A)$ is nonlinear, it forces us to reconsider how we should really generalize the Jacobson's method.} can be obtained and so in principle there should not be a multi-center problem (This remains to be seen explicitly since \cite{2602.20430} did not develop the multi-horizon case). 

There is yet another curious analogy with the historical development of astronomy. A prescription of the form $G_\text{eff}(r)$ naturally privileges a center, reminiscent of the geometrical constructions of Ptolemaic astronomy. Indeed, the Ptolemaic system even introduced the equant, a distinguished point displaced from the geometric center of a deferent, with respect to which the angular motion was uniform. Heliocentrism replaced the Earth by the Sun as the organizing center of planetary motion, but this alone does not answer a more general question: if the Sun is \emph{the} center, why does an apple fall toward the Earth rather than toward the Sun? Newtonian gravity ultimately removes the question rather than choosing yet another center: every mass acts as a gravitational source, and there is no distinguished gravitational center of the Universe. An analogous distinction arises here. For a single isolated black hole, $G_\text{eff}(r)$ causes no difficulty. In a multi-center configuration, however, an observation point $x$ is associated with several distances $r_i=|x-x_i|$, and hence potentially several source-dependent values $G_\text{eff}(r_i)$. As discussed above, this is not in itself an observational inconsistency: the measurable combinations are $G_{\rm eff}(r_i)M_i$, which an observer assuming constant $G$ could simply interpret as different effective masses. The deeper question is whether the single-center prescription supplies a consistent rule for combining these responses beyond the linear regime. A field-theoretic $G(x)$ equipped with its own EOM, or a nonlocal kernel $G(x,y)$ in the yet-to-be-constructed full theory, provides possible center-independent completions. The issue is not that multiple centers are forbidden, but that a radial prescription alone does not determine how the corresponding responses are to be reconciled dynamically. GEVAG avoids this particular issue at the horizon level by \emph{not} promoting $G_\text{eff}$ to a radial bulk function in the first place, which is conceptually closer to removing the question of a preferred center from ever arising. However, even then, the gravitational susceptibility interpretation of GEVAG \cite{companion}, i.e. Eq.(\ref{sus}), is only strictly true when the response is linear. So again, the full nonlinear theory requires further studies in GEVAG even though it has no multi-center problem. 

Not only that generalized entropy does not uniquely determine the geometry, it also does not define ``the'' effective Newton constant until one specifies what physical question defines it. On the theoretical side, we have already known from our experience with many modified gravity theories that in general, the gravitational coupling appearing in the background equations, the perturbation equations, local force law, lensing potential, gravitational wave, matter sector, etc. need not be identical. In the generalized entropy case, there is also no reason to suspect that they do. Perhaps GEVAG, EGC, the Wald reconstruction, Cavendish measurements, etc. are different operational questions. The natural susceptibility language for GEVAG suggests it is a gravitational response $G_\text{resp}$; the EGC construct relates it to global geometry so it is more like $G_\text{geom}$; the Wald entropy approach essentially relates it with the coefficient associated with the gravitational Lagrangian and we have a distinct $G_\text{Wald}$; and finally the experimental Cavendish measurements detect another $G_\text{Cav}$. There is no \emph{a priori} reason why these need to be the same. GR is exceptional precisely because the linear area law gives the same $G$ for all these questions. This is similar to the equivalence between inertial mass and gravitational mass in GR.
Perhaps Nature chooses the simplest theory after all, in the sense that various inequivalent notions become equivalent.

\begin{quote}
``We are to admit no more causes of natural things than such as are both true and sufficient to explain their appearances. To this purpose the philosophers say that Nature does nothing in vain, and more is in vain when less will serve; for Nature is pleased with simplicity and affects not the pomp of superfluous causes.'' --- Isaac Newton, \emph{Principia} (1687)
\end{quote}



\begin{acknowledgments}
This research is supported by NUAA funding No.1018-ILF26028.
\end{acknowledgments}

\end{document}